# Theory of Topological Corner State Laser in Kagome Waveguide Arrays

Hua Zhong[1,2], Yaroslav V. Kartashov,[3,4] Alexander Szameit[5], Yongdong Li,[1] Chunliang Liu,[1] and Yiqi Zhang[1,2,*]

[1]*Key Laboratory for Physical Electronics and Devices of the Ministry of Education & Shaanxi Key Lab of Information Photonic Technique, School of Electronic and Information Engineering, Xi'an Jiaotong University, Xi'an 710049, China*

[2]*Guangdong Xi'an Jiaotong University Academy, Foshan 528300, China*

[3]*ICFO-Institut de Ciències Fotòniques, The Barcelona Institute of Science and Technology, 08860 Castelldefels (Barcelona), Spain*

[4]*Institute of Spectroscopy, Russian Academy of Sciences, Troitsk, Moscow, 108840, Russia*

[5]*Institute for Physics, University of Rostock, Rostock, 18059 Rostock, Germany*

[*]*Corresponding author: zhangyiqi@mail.xjtu.edu.cn*

**Abstract:** In comparison with conventional lasers, topological lasers are more robust and can be immune to disorder or defects if lasing occurs in topologically protected states. Previously reported topological lasers were almost exclusively based on the first-order photonic topological insulators. Here, we show that lasing can be achieved in the zero-dimensional corner state in a second-order photonic topological insulator, which is based on Kagome waveguide array with a rhombic configuration. If gain is present in the corner of the structure, where topological corner state resides, stable lasing in this state is achieved, with lowest possible threshold, in the presence of uniform losses and two-photon absorption. When gain acts in other corners of the structure, lasing may occur in edge or bulk states, but it requires substantially larger thresholds and transition to stable lasing occurs over much larger propagation distances, sometimes due to instabilities, which are absent for lasing in corner states. We find that increasing two-photon absorption generally plays strong stabilizing action for nonlinear lasing states. The transition to stable lasing stimulated by noisy inputs is illustrated. Our work demonstrates the realistic setting for corner state laser based on higher-order topological insulator realised with waveguide arrays.

## 1 Introduction

Generally, a $d$-dimensional ($d$D) topological insulator supports ($d$D) bulk states and ($d-\ell$)D topological edge states. First-order TIs correspond to $\ell=1$, while for higher-order topological insulators (HOTIs) one has $\ell>1$ [1,2]. Nowadays, HOTIs attract considerable attention in diverse areas of physics, as they were discovered in condensed matter physics [3-10], electrical systems [11], mechanical systems [12,13], acoustics [14-17], microwave systems [18], and photonics [19-26]. Different from first-order topological insulators that satisfy the bulk-edge correspondence principle [27,28], HOTIs do not comply with this principle, even though they support topologically protected states [14,15]. The simplest realization of a HOTI is a 2D insulator supporting 0D topological corner states. As $\ell=2$ such a system is called second-order topological insulator HOTIs were reported not only in conservative systems, but also in non-Hermitian settings [29-32]. Moreover, only recently nonlinear HOTIs were considered theoretically, with nonlinearity-dependent hopping rates that enters topological phase at high enough amplitudes [33], and coupling between exciton-polariton corner modes [34].

Nonlinear effects are imperative for implementing laser systems. Nonlinearity leads to competition between different lasing modes, as a result of which only limited set of them survives, leading to substantial modification of the output spectrum in stable lasing regime. Gain saturation, accounted for by nonlinear terms in corresponding governing equations, determines the amplitudes of lasing modes. These effects were employed for the realization of so-called topological lasers [35-41] representing a novel extension of the concept of topological insulators. In comparison with conventional lasers, whose stability may be affected by perturbations, such as defects and disorder, topological lasers may be more stable when they lase in topologically protected edge states. Topological lasing has been theoretically proposed and experimentally reported in the 1D systems representing variants of the Su-Schrieffer-Heeger (SSH) lattices [35-38], and in fully 2D settings, based on topological photonic crystals [42] and lattices of coupled-ring resonators [43,44]. Lasers based on the valley Hall effect were reported in [45-48]. Topological lasing is possible in Floquet topological insulators [49] and it was proposed in bosonic Harper-Hofstadter model [50]. In polariton systems, topological lasing in the 1D and 2D microresonator arrays have been demonstrated experimentally [36,51], and studied theoretically when sufficiently strong pump and, in some cases external magnetic field, were provided in the system [52].

Remarkably, in all previous works, even those addressing finite topological configurations [53], lasing in 2D systems was shown to occur in the extended edge states, typically along the entire periphery of the insulator [42-44]. In contrast, HOTIs potentially allow the realization of a class of higher-order topological lasers, where despite the fact that the system will still show fully two-dimensional evolution, lasing will occur in localized 0D corner modes enabling high spatial localization of the emission and its constant modal population (in a sense that only one nonlinear mode will be excited above the lasing threshold). In addition, corner lasers constructed on topologically protected edge states should exhibit benefits of topological protection of lasing. This is because corner states always survive even for harsh perturbations in the bulk of the photonic lattice, because topological properties are not destroyed unless perturbation-induced shifts of propagation constants exceed the width of the topological gap.

The goal of this work is to demonstrate that corner laser can be implemented in a new continuous system, which is based on Kagome [14,15,22,54] waveguide arrays, where localized gain can be selectively provided in different waveguides of the array, and where uniform losses, two-photon absorption, and focusing nonlinear interactions are present. Even though lasing in 0D corner states has been recently reported in photonic crystal cavities [55-57], and somewhat similar effect resulting in lasing in vertices of triangular valley-Hall laser with controllable degree of asymmetry in the underlying periodic structure was reported in [58], our work is different in many aspects. We provide the illustration of topological corner lasers on new platform – shallow waveguide arrays (as opposed to previous works on microrings or photonic crystal cavities with 2D SSH configuration) – the idea that can be extended to polaritonic systems based on micropillar arrays, where nonlinear interactions of polaritons are repulsive [36,51,52] (in our case nonlinearity is attractive). Continuous model, employed here, takes into account all features of the refractive index and gain landscapes, as opposed to simplified discrete models of HOTI lasers introduced before. We employ localized gain, which, when applied in different corners of our structure, offers unique advantage of highly selective excitation of bulk, edge, or corner nonlinear topological states in HOTIs. The possibility of such selective excitation in our truly two-dimensional system is a nontrivial result by itself taking into account complex spatial shape of the considered structure. We report on completely stable corner state lasing and bistability of the edge state lasing in higher-order topological insulators that can only be observed in the nonlinear medium. The detailed analysis of stability of lasing modes is provided for different values of linear gain and two-photon absorption coefficients. It is illustrated how lasing in corner states builds up from random noisy inputs. Stability of lasing in the presence of disorder is illustrated too.

Even when 0D corner states supported by our structure are strongly localized, they always penetrate into neighboring waveguides, where field typically changes its sign. Importantly, by tuning the separation between waveguides, one can substantially change the area of the topological corner modes, that is not achievable, say, with single-element structures. The above mentioned tunability of corner mode area, in the case when lasing occurs in such a mode, may be used for the enhancement of the light-matter interactions and in the design of nonlinear photonic devices with better characteristics, as compared to usual topological lasers, where lasing is usually achieved in extended modes occupying the entire periphery of the structure. Also, in corner laser the appearance of topological modes is guaranteed by simple deformation of the structure allowing miniaturization and increasing performance in the nonlinear regime.

## 2 Results and Discussion

### 2.1 Band structure and linear modes of the conservative system

The propagation dynamics of light beams in our dissipative system – Kagome array of waveguides with tunable spacing – can be described by the nonlinear Schrödinger-like equation that in the dimensionless units reads as:

$$i\frac{\partial \psi}{\partial z} = -\frac{1}{2}\left(\frac{\partial^2}{\partial x^2} + \frac{\partial^2}{\partial y^2}\right)\psi - (\mathcal{R}_{\text{re}} - i\mathcal{R}_{\text{im}} + i\gamma)\psi - (1 + i\alpha)|\psi|^2\psi. \tag{1}$$

Here $\psi = (\kappa^2 w^2 n_{2,\text{re}}/n_{\text{re}})^{1/2} E$ is the dimensionless field amplitude; $x, y$ are the transverse coordinates normalized to the characteristic scale $w$; $z$ is the propagation distance scaled to the diffraction length $\kappa w^2$; $\kappa = 2\pi n_{\text{re}}/\lambda$ is the wavenumber; $n_{\text{re}}$ and $n_{\text{im}}$ ($n_{\text{im}} \ll n_{\text{re}}$) are the real and imaginary parts of the unperturbed linear refractive index of the material, respectively; $n_{2,\text{re}}$ and $n_{2,\text{im}}$ are the real and imaginary parts of the nonlinear refractive index, respectively; $\gamma = \kappa^2 w^2 n_{\text{im}}/n_{\text{re}}$ is the coefficient of linear losses that are assumed uniform; $\alpha = n_{2,\text{im}}/n_{2,\text{re}}$ is the scaled coefficient characterizing nonlinear losses stemming from all sources, including intrinsic nonlinear losses of the medium and gain saturation in the first approximation. The refractive index distribution is described by the function $\mathcal{R}_{\text{re}}(x, y) = p_{\text{re}} \sum_{n,m} \mathcal{Q}(x - x_n, y - y_m)$ that is composed of Gaussian waveguides $\mathcal{Q} = \exp[-(x^2 + y^2)/d^2]$ with normalized depths of $p_{\text{re}} = \kappa^2 w^2 \delta n_{\text{re}}/n_{\text{re}}$, where $(x_n, y_m)$ are the coordinates of the sites of the Kagome lattice and $d = 0.5$ the waveguide width. Here we choose $p_{\text{re}} = 10$ to make sure that every channel is a single-mode waveguide. For definition of $\mathcal{R}_{\text{im}}$ see section 2.3 below.

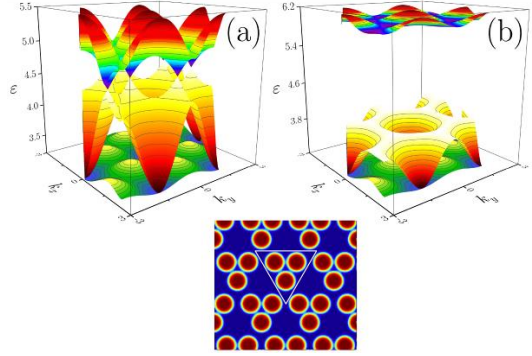

Fig.1. First three bands for arrays with $r = 0.5a$ (a) and $r = 0.4a$ (b). The inset between panels shows array profile at $r = 0.55a$. White triangle illustrates unit cell of the array.

Kagome waveguide arrays considered here can be deformed in a controllable fashion upon fabrication. The controllable deformation is described by the parameter $r$ defining shift of each second waveguide in the structure: see Fig. 1 for the band structure of the extended Kagome array and its profile with the unit cell indicated in the inset (also see middle-left panel in Fig. 2, where shift $r$ and spacing $a = 2.5$ are indicated on the array profile shown with white contour lines). When $r = a/2$ one retrieves the standard Kagome array, and the band structure exhibits 6 Dirac cones between the upper two bands [Fig. 1(a)]. When $r < a/2$, the band gap opens, as shown in Fig. 1(b).

We further truncate infinite Kagome array such, that finite rhombic configuration forms (see white contours in panels of Fig. 2). The resulting system is essentially two-dimensional, since it is obtained by truncation of truly two-dimensional Kagome array and it is not equivalent to folded one-dimensional SSH chains. As we show below, our system possesses a unique spectrum of eigenmodes, where both bulk, edge, and corner modes are present in proper parameter range. Corner states in this truncated structure appear when the structure is topologically nontrivial at $r < a/2$. For the rhombic configuration in Fig. 2, they form in the top right corner of the array. Due to specific duality of our structure (that for the case of $r > a/2$ is equivalent to the structure with $r < a/2$ rotated by an angle of $\pi$), for the case with $r > a/2$, such states also emerge in the bottom left corner. Notice that this duality is exclusively due to our choice of truncation of the structure (see below).

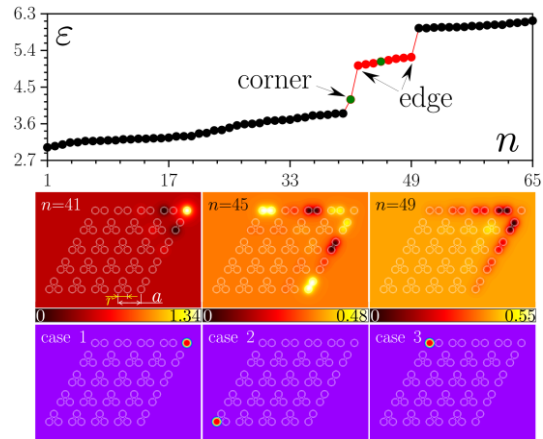

Fig. 2. Spectrum, corner and edge states of the conservative system. Top row: Linear spectrum $\varepsilon(n)$ of the array at $p_{\rm im} = 0$, where n is the eigenvalue number. Color dots correspond to corner and edge states, while black dots correspond to bulk modes. Green dots indicate modes that lase first in the presence of localized gain acting in the top-right or top-left sites of the array. Middle row: Representative profiles of the corner and edge modes supported by the structure at $p_{\rm im} = 0$. Array sites are indicated with white lines. Bottom row: Three different gain configurations that we consider, with gain profile indicated by the red spot superimposed on the array profile.

First, we consider the spectrum of the linear modes of our truncated array without gain and losses ($p_{\rm im}, \gamma = 0$) and by omitting the last, nonlinear term in the Eq. (1). We find such modes from Eq. (1) in the form $\psi(x, y, z) = u(x, y)\exp(i\varepsilon z)$ by solving the resulting linear eigenvalue problem $\varepsilon u = (1/2)(\partial_x^2 + \partial_y^2)u + \mathcal{R}_{\rm re} u$. For the array with $r = 0.4a$, the energies $\varepsilon$ (propagation constants) of modes are shown in the top panel in Fig. 2. This spectrum includes one 0D corner state with eigenvalue number $n = 41$, a band of 1D edge states at $42 \leq n \leq 49$ shown by the red dots, and bulk modes (black dots). This structure of spectrum remains qualitatively similar for larger arrays. In the middle row of Fig. 2 we show the real-valued amplitude distributions $u(x,y)$ for the conservative corner ($n = 41$) and edge states at $n = 45$ (green dot) and $n = 49$ (edge state with largest $\varepsilon$). Notice strong localization of light in the topological mode in the top-right corner of our continuous structure [14,15,22] for selected $r$ value. Even in this case, however, the mode penetrates into neighboring waveguides, where field changes its sign. In fact, localization of corner modes strongly depends on the difference $r - a/2$ and such modes may extend far beyond corner waveguide, when the above difference is small, demonstrating again two-dimensional nature of our system. This possibility to control localization of topological corner modes will be also very important in nonlinear lasing regime considered below, where by tuning $r - a/2$ one may control the area of the lasing mode, in contrast with the case of single waveguide, for example. Edge states that also appear in our system are characterized by excitations residing in the vicinity of two edges adjacent to the top right corner and they differ only by the number of nodes (where $\psi$ vanishes) in the state along these edges.

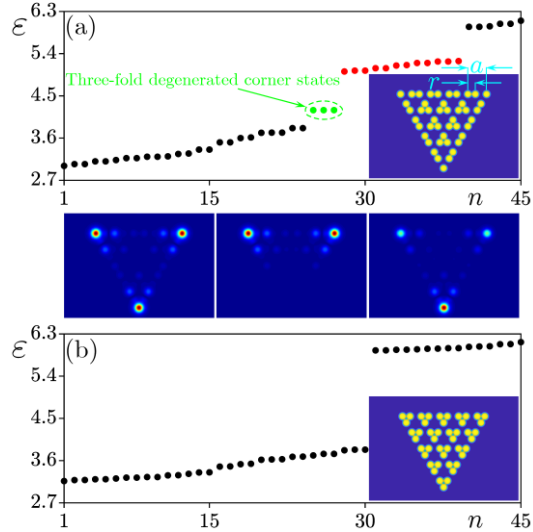

Fig. 3. Linear spectrum of modes in the triangular configuration. (a) $r = 0.4a$. Green dots are the corner states, red dots are the edge states, while black dots are bulk states. The bottom panels show the three degenerated corner states. (b) $r = 0.6a$. There are only bulk states.

In contrast to the rhombic configuration, the above mentioned duality is absent in the triangular Kagome configuration obtained from the same infinite Kagome array (see Fig. 3). Therefore, upon modifications of the parameter $r$ controlling the shift of waveguides, topological corner states in triangular structure emerge *only* in the regime, where $r < a/2$. To prove this, in Fig. 3(a) we show the spectrum of linear modes in the triangular configuration with $r = 0.4a$. The waveguide array with triangular shape is shown in the inset. Due to the symmetry of the configuration, corner states simultaneously emerge in all three equivalent corners of the structure (green dots) and they have nearly identical propagation constants. Their field modulus distributions are shown in the panels below the spectrum. Notice that these corner states are linear, so an arbitrary linear superposition of such corner states also generates a correct corner state of this system. The red dots in the spectrum correspond to the edge states localized on the sides of the triangle. To prove that in this structure corner states emerge only at $r < a/2$, in Fig. 3(b) we show the linear spectrum of this triangular structure at $r = 0.6a$ (see also Fig. S7 in **Supplemental Materials**). There are no corner or edge states in the spectrum. This case is therefore topologically trivial. It should be stressed that the energies of corner states at $r < a/2$ in the triangular configuration are the same as energy of the corner state in rhombic configuration in Fig. 2. The only difference is that in triangular configuration corner states are triply degenerate and appear simultaneously in all three equivalent corners of the structure. Consequently, in the presence of gain and losses lasing thresholds for corner states in rhombic and triangular configurations are also identical, as further discussed in **Supplemental Materials**. Since there is no degeneracy of corner states in the rhombic configuration, further throughout this paper we adopt this structure for the analysis of the HOTI laser.

It should be mentioned that the topological properties of the Kagome array can be characterized by the bulk polarization that can be calculated using the formula $p_j = -S^{-1} \iint A_j d^2 k$, where $A_j = -i\langle u | \partial k_j | u \rangle$ is the Berry connection, and $S$ is the area of the first Brillouin zone (BZ), $u$ is the Bloch mode of the structure. The details of calculation of the bulk polarization are presented in the **Supplemental Materials**. The eigenvectors $u$ can be calculated using the tight-binding Hamiltonian of the system. The system is in topological phase when the polarization components are nonzero and in trivial phase when the polarization is zero. The bulk polarizations $(p_x', p_y')$ are calculated in the transformed coordinate system, where BZ is square, with $(p_x', p_y') = (0,0)$ for $r > a/2$ (topologically trivial phase) and $(p_x', p_y') = (1/3, 1/3)$ for $r < a/2$ (topologically nontrivial phase). Thus, in our case, the appearance of corner states in the spectrum of continuous system upon variation of $r$ was always correlated with the emergence of the nontrivial bulk polarization. Thus, further we consider representative array with rhombic configuration and $r = 0.4a$ illustrated in Fig. 2.

## 2.2 Linear modes of the dissipative system

To realize lasing, we provide localized gain in one of the corner sites of our rhombic structure (labelled as case 1, case 2, and case 3, respectively, in the bottom row of Fig. 2). The inhomogeneous gain landscape is described in Eq. (1) by the function $\mathcal{R}_{\text{im}}(x, y) =$ $p_{im}Q(x - x_c, y - y_c)$, where $x_c, y_c$ are the coordinates of corresponding corner waveguide, while normalized gain amplitude is given by $p_{im} = \kappa^2 w^2 \delta n_{im}/n_{re}$ ($p_{im} \ll p_{re}$).

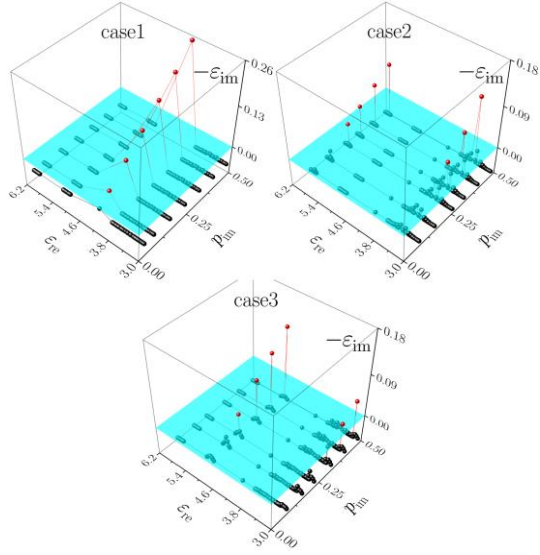

Fig. 4. Energy $\varepsilon = \varepsilon_{re} + i\varepsilon_{im}$ versus gain amplitude $p_{im}$ for all modes of the system for three different locations of the channel with gain shown in Fig. 2. The real part $\varepsilon_{re}$ of energy and its imaginary part $-\varepsilon_{im}$ with inverted sign are shown simultaneously on two different axes, since they may simultaneously change upon variation of the gain amplitude $p_{im}$. For selected $p_{im}$ the projection of each dot on the $(\varepsilon_{re}, -\varepsilon_{im})$ plane gives the energy of corresponding mode in the linear case. Only those modes that have $-\varepsilon_{im} > 0$ may lase. The dots corresponding to each lasing mode are shown red and they are located above the semi-transparent cyan plane corresponding to $\varepsilon_{im} = 0$. From this plot one can see the modes from which group of bands start lasing first.

To understand how inhomogeneous gain and uniform losses affect linear modes of this structure we calculate the spectrum of the system still neglecting the nonlinear effects, but setting $\gamma = 0.05$ and increasing $p_{im}$. The corresponding complex eigenvalue problem can be written as $\varepsilon u = (1/2)(\partial_x^2 + \partial_y^2)u + (\mathcal{R}_{re} - i\mathcal{R}_{im} + i\gamma)u$, where $\varepsilon = \varepsilon_{re} + i\varepsilon_{im}$, $\varepsilon_{re}$ and $\varepsilon_{im}$ are the real and imaginary parts of eigenvalues, respectively. The sign of the imaginary part $\varepsilon_{im}$ is determined by the losses $\gamma$ and gain $\mathcal{R}_{im}$. If $\varepsilon_{im} < 0$, the modes are amplified, if $\varepsilon_{im} > 0$ they are damped. Figure 4 shows how $\varepsilon_{re}$ and $\varepsilon_{im}$ vary with increase of the gain amplitude $p_{im}$ for three different gain positions, corresponding to the cases 1, 2, and 3, outlined in Fig. 2. In Fig. 4 we plot the inverted value $-\varepsilon_{im}$ for illustrative purposes. Positive values of $\varepsilon_{im}$ shown by black dots and lying below the cyan plane ($\varepsilon_{im} \equiv 0$) are thus associated with damped modes, while negative $\varepsilon_{im}$ values corresponding to the red dots lying above the cyan plane are associated with amplified states. Real parts $\varepsilon_{re}$ are weakly affected by gain amplitude $p_{im}$. One can see, that in all three cases growing modes appear when gain amplitude $p_{im}$ exceeds certain threshold, but there are important differences between these gain arrangements. For case 1, when gain is located in the corner supporting topological corner state, only this state is amplified in the broad range of $p_{im}$ values, while all other states are damped. Numerical simulations demonstrate that the lasing threshold for the topological corner state in case 1 takes lowest possible value $p_{im}^{th} \sim 0.078$ for this structure. The lasing threshold remains the same in the triangular configuration (see **Supplemental Materials**). For the case 2, only bulk states can lase above the highest lasing threshold $p_{im}^{th} \sim 0.255$. The number of such states increases with increase of $p_{im}$. In the case 3, the edge state with $n = 45$ (whose counterpart in conservative system is depicted in Fig. 2) lases first above the threshold $p_{im}^{th} \sim 0.225$, but further increase of gain amplitude results in amplification of some of the bulk modes. Corner state is always damped in the cases 2 and 3 in the interval of $p_{im}$ values considered here, so one can expect to realize topological lasing only in the case 1. This is natural taking into account that amplification efficiency is determined by the overlap of the corner mode with gain landscape, which is highest in the latter case, that also explains lowest lasing threshold.

## 2.3 Families of nonlinear lasing modes

Amplification of the corner and edge states at $p_{im} > p_{im}^{th}$ can be eventually arrested by the nonlinear absorption. To explore the possibility of the exact and stable balance between diffraction, nonlinearity, gain, and absorption in this system we now consider complete model (1) with all nonlinear terms included and search for stationary nonlinear corner and edge states with constant power along propagation distance. Their profiles are described by the equation

$$\varepsilon u = \frac{1}{2}\left(\frac{\partial^2}{\partial x^2} + \frac{\partial^2}{\partial y^2}\right)u + (\mathcal{R}_{re} - i\mathcal{R}_{im} + i\gamma)u + (1 + i\alpha)|u|^2 u, \qquad (2)$$

with the real-valued nonlinear energy shift (or propagation constant) $\varepsilon$ determined by the Kerr nonlinearity and two-photon absorption. On physical grounds, stabilization of the lasing modes is achieved due to two-photon absorption that prevents the amplitude from unlimited growth above the lasing threshold. Nonlinear absorption counteracts the tendency for self-localization due to self-focusing nonlinearity, it leads to the appearance of nontrivial internal currents in nonlinear solution that are directed outward the pump spot, and it results in spatial broadening of stable lasing states for larger values of $\alpha$ coefficient. Since $\varepsilon$ is not an independent parameter in this dissipative system, we use Newton method complemented by the power balance condition to obtain families of the nonlinear lasing states parameterized by the gain amplitude $p_{\text{im}}$, that can be written as:

$$\iint [(\mathcal{R}_{\text{im}} - \gamma)|\psi|^2 - \alpha|\psi|^4]dxdy = 0 \tag{3}$$

Notice that standard simulation of evolution does not provide unstable branches, because in such a method the wave converges only to dynamically stable attractors. In contrast, our method, based on simultaneous solution of Eqs. (2) and (3) allows to get all solutions which may be stable or unstable, and even determine bistable regimes.

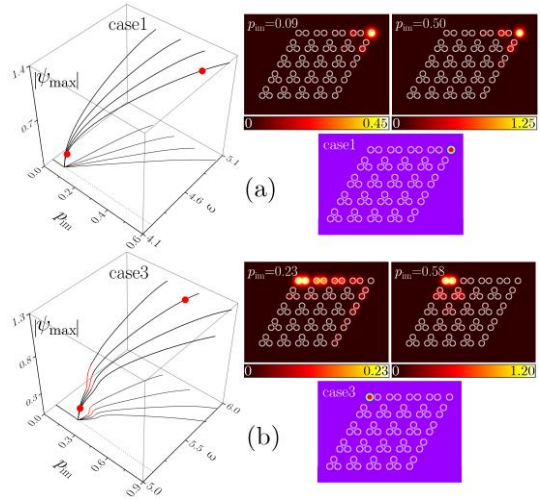

Fig. 5. Nonlinear lasing state families. Maximal amplitude $\psi_{\text{max}}$ and propagation constant $\varepsilon$ of the nonlinear lasing mode versus gain amplitude $p_{\text{im}}$ for two different locations of the amplifying channel [(a) – case 1, (b) – case 3] and nonlinear absorption coefficients $\alpha = 0.1, 0.2, 0.3,$ and $0.4$ (in all plots left outermost curve corresponds to $\alpha = 0.1$, while right one to $\alpha = 0.4$). Stable branches are shown black, unstable branches are shown red. Red dots correspond to $|\psi(x, y)|$ distributions shown in the top row. Bottom row schematically shows channel with gain in the array.

To characterize nonlinear families of the corner and edge states (emerging in the cases 1 and 3, respectively) in our system, we plot their peak amplitude $|\psi|_{\text{max}}$ and nonlinear energy shifts $\varepsilon$ as functions of $p_{\text{im}}$ in the left panels of Figs. 5(a) and 5(b), respectively, for different nonlinear absorption coefficients $\alpha$. Corresponding gain landscapes are presented for each case too. In both cases stationary nonlinear modes appear when gain amplitude $p_{\text{im}}$ exceeds corresponding lasing threshold, $p_{\text{im}}^{\text{th}} \sim 0.078$ in the case 1. These states are characterized by complex internal currents, but for them nonlinear absorption exactly integrally compensates inhomogeneous gain. Nonlinear energy shift $\varepsilon$ increases almost linearly with $p_{\text{im}}$ in the region $p_{\text{im}} > p_{\text{im}}^{\text{th}}$ (all modes are damped in the region $p_{\text{im}} < p_{\text{im}}^{\text{th}}$). At $p_{\text{im}} = p_{\text{im}}^{\text{th}}$ it naturally coincides with real part of the eigenvalue $\varepsilon_{\text{re}}$ of the linear edge state from which nonlinear mode bifurcates [as shown by the bottom dotted line in the $(p_{\text{im}}, \varepsilon)$ plane in Fig. 5(a)], allowing to uniquely identify the state giving rise to the stable lasing mode. Thus, such nonlinear corner states are of topological origin since they bifurcate from 0D linear topological corner modes [1,14,15,22]. Since we only consider corner lasing modes forming in the gap that do not couple to edge or bulk modes, we show in Fig. 5(a) the curves for $\varepsilon$ values below the band occupied by the eigenvalues of the linear edge states (i.e. below red dots in the spectrum of Fig. 2). In the top two panels in the right column in Fig. 5(a) we show $|\psi|$ distributions for corner lasing states at $\alpha = 0.4$ for different gain amplitudes, corresponding to the red dots, one close to lasing threshold ($p_{\text{im}} = 0.09$) and one rather far from it ($p_{\text{im}} = 0.50$). One can see that for $r = 0.4a$ such states are very well localized and that most of their energy is concentrated in the top-right corner site. However, when $r \to a/2$ lasing can be achieved even in weakly localized corner modes confirming tunability of this system mentioned in the introduction. Quantitatively similar results are also obtained in the triangular configuration, as shown in the **Supplemental Materials**.

In contrast, when gain is provided in the top left corner of the array, that corresponds to the case 3 [see the bottom panels in the right column of Fig. 5(b)], the conditions for the most efficient amplification are met for the edge state with $n = 45$ [upper green dot in Fig. 2] that starts lasing when gain amplitude exceeds threshold $p_{\text{im}}^{\text{th}} \sim 0.225$ that nevertheless is substantially larger than the threshold obtained for the corner state in the case 1. Still, it is remarkable that by choosing gain location in this system one can selectively excite either 0D corner or 1D edge states. This illustrates a rich variety of lasing regimes that can be observed in our structure that would not be available in single-waveguide geometries, for example. Nonlinear families corresponding to different absorption coefficients for the case 3 are displayed in the left panel of Fig. 5(b). Nonlinear lasing edge state bifurcates from the linear edge state with $n = 45$, its energy grows with $p_{\rm im}$ until it reaches the gap edge, above which coupling with bulk modes occurs. Interestingly, despite the fact that usual linear amplification is used, rather than pump with certain energy, the bistability domain is encountered for edge states in the case 3, where three solutions can coexist for the same $p_{\rm im}$ value. The encountered bistability offers the unique advantage to achieve lasing in the edge states with different internal structures and amplitudes for certain $p_{\rm im}$ values. $|\psi|$ distributions from the middle branch of $\varepsilon(p_{\rm im})$ dependence clearly show strong coupling with the opposite edge of the rhombic structure. In the top panels in the right column of Fig. 5(b), we illustrate representative $|\psi|$ distributions from the lower and upper branches corresponding to the red dots at $p_{\rm im} = 0.23$ and $p_{\rm im} = 0.58$. With increasing gain amplitude $p_{\rm im}$, initially extended over two adjacent sides of the rhombus edge state gradually contracts toward top left corner of the structure, so that the profile of the nonlinear edge lasing state may strongly differ from its linear counterpart shown in the middle panel in the second row ($n = 45$) in Fig. 2. From the dependencies shown in Fig. 5 one can see that for a fixed gain amplitude $p_{\rm im}$, the peak amplitude of the nonlinear lasing state decreases with increase of the nonlinear absorption coefficient $\alpha$. The energy interval, where nonlinear states exist increases with increase of the nonlinear absorption.

On physical grounds, one can perform comparison of lasing in corner states of Kagome array and in isolated waveguide. While there is no fundamental difference in lasing efficiency, Kagome array provides tunability that is absent for single waveguide, since the structure of the lasing mode in Kagome array strongly depends on the dimerization parameter $r$. The field of corner lasing mode is localized mostly on one of the sublattices forming Kagome array and it changes its sign between unit cells, that results in staggered tails, as it is obvious from profile of the corner mode ($n = 41$) from Fig. 2. Thus, the presence of the array does allow to control the spatial extent of the lasing pattern and is therefore very important (thus, at $r$ values close to $a/2$ lasing will occur in the mode strongly penetrating into the depth of array, even though gain is provided only in one waveguide). A detailed comparison between lasing in Kagome array and in the isolated waveguide is presented in the **Supplemental Materials**.

Stability of the nonlinear lasing modes is central from the point of view of construction of topological laser, because it guarantees that the only spatial mode will be excited and there will be no oscillations due to instabilities or beatings between several excited modes. Stability analysis for nonlinear lasing modes obtained with Newton method was performed by adding into their profiles small-amplitude noise (typically up to 1% in amplitude) and propagating them (by using split-step Fourier method) over huge distances $z \sim 10^4$, far exceeding the length of any realistic sample. In all cases the amplitude of modes, which are stable, was returning to the unperturbed value and such modes kept their internal structure. This method allows capturing even weak instabilities and accurately determining stability domains. The regions where lasing is stable or unstable are identified in Fig. 5, where black curves correspond to stable lasing states, while red ones to unstable states. We found that corner lasing states in the case 1 are always stable, even for low nonlinear absorption coefficients [Fig. 5(a)]. Nonlinear edge states in the case 3 can be stable too, but the middle branches in the encountered bistability regions are always unstable. The bistability illustrated in Fig. 5(b) is a result of nonlinear competition between several coexisting edge states that all experience amplification. Indeed, we have found that the structure of the lasing state substantially changes in the red unstable region in Fig. 5(b) with increase of its peak amplitude, i.e. at low and high peak amplitudes the relative weights of different edge states in lasing mode change because they experience different gain (the overlap of their spatial profiles with pump spot is different) and at the same time relative weights of modes determine total absorption experienced by the wave. Due to this competition several stationary nonlinear configurations can coexist for the same gain amplitude, giving rise to bistability. The unstable regions gradually disappear with increase of the nonlinear absorption, until the entire branch of the edge states becomes stable [Fig. 5(b)].

## 2.4 Noise-stimulated lasing

It is interesting to mention that lasing states in this dissipative system, being stable attractors, can be excited from the broad range of the initial conditions. For instance, they emerge from broadband random noise that initially excites all modes of the system, upon subsequent nonlinear competition between these modes, some of which exhibit preferential amplification. This once again confirms robustness of the nonlinear lasing states. In order to illustrate this type of laser operation, in Fig. 6 we show different stages of the evolution dynamics of noisy input excitations for gain landscapes corresponding to the cases 1 and 2. Figure 6(a) shows peak amplitude $|\psi|_{\rm max}$ of the field versus propagation distance $z$ for the case 2. One can see that peak amplitude $|\psi|_{\rm max}$ exhibits long gradually decaying oscillations approaching the stationary value only after a long propagation distance $z \sim 750$. Corresponding $|\psi|$ distributions reveal transition from noisy input to regular, but extremely extended pattern that occupies practically the entire waveguide array. Even though this pattern features somewhat larger amplitude in the bottom left corner due to high localization of gain profile, it clearly indicates excitation of the bulk modes of the structure in accordance with spectrum of Fig. 4 (increasing the size of the waveguide array leads to further expansion of this pattern). In contrast, for gain landscape corresponding to the case 1, the peak amplitude of the excitation quickly reaches (at the distance $z \sim 100$) its asymptotic value, as illustrated in Fig. 6(b). The nonlinear mode quickly concentrates in the top right corner, supporting topological corner state, and then remains strongly localized and stable at all propagation distances.

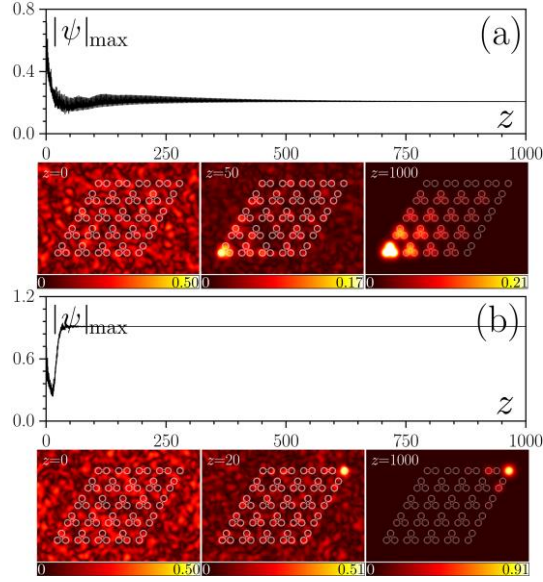

Fig. 6. Lasing states excited by noisy inputs. Excitation dynamics of nonlinear lasing modes at $p_{\text{im}} = 0.3$ and $\alpha = 0.4$ in case 2, when gain is provided in the corner that does not support topological state (a) and case 1, when gain acts in the corner where topological mode resides (b), from noisy inputs. In each case the dependence of peak amplitude $|\psi|_{\text{max}}$ on distance is shown together with $|\psi|$ distributions at selected propagation distances.

## 2.5 Discussion

Propagation of light in shallow waveguide arrays adopted in this work is perfectly described within the frames of the paraxial approximation assuming excitation at a single frequency, where appropriate spatially-inhomogeneous gain can be realized with existing techniques. Nonlinear dissipative states emerging in such arrays are the result of competition between different spatial modes of the system, some of which are preferentially amplified, but all nonlinear dissipative states reported here can be traced back to the linear modes of the structure from which they emerge at threshold gain amplitude $p_{\text{im}} = p_{\text{im}}^{\text{th}}$, the fact that allows to claim single spatial mode lasing in this system. The competition between modes occurs due to conservative and dissipative parts of nonlinearity present in the Eq. (1), that determines also dynamical stability of the emerging states. The results reported here can be extended to other optical or optoelectronic systems, such as photonic crystals, photonic crystal fibers and structured polariton microcavities, where spatially-inhomogeneous gain and suitable refractive index/potential energy landscapes can be fabricated or induced optically.

The system based on the waveguide arrays suggested in this work can be practically implemented with doped (for example, with Er) chalcogenide glasses (such as GaLaS or AsSe), where waveguide arrays can be written with tightly focused laser pulses. The nonlinear coefficient in such materials is about $n_{2,\text{re}} \sim 1 \times 10^{-17}$ m$^2$/W, while nonlinear absorption coefficient varies between $n_{2,\text{im}} \sim 10^{-19} - 10^{-17}$ m$^2$/W depending on the composition of the glass [59-61]. In this case, one finds that when transverse coordinates are scaled to 10 $\mu$m, the dimensionless propagation distance in Eq. (1) is scaled to the diffraction length of $\sim 1.5$ mm; the waveguide depth of $p_{\text{re}} = 10$ corresponds to real refractive index contrast $\sim 1.13 \times 10^{-3}$; while gain amplitude $p_{\text{im}} = 0.1$ corresponds to $\delta n_{\text{im}} \sim 1 \times 10^{-5}$ for unperturbed refractive index $n_{\text{re}} \sim 2.81$ at the wavelength $\lambda = 1.08$ $\mu$m. The structure proposed here can also be implemented in planar pumped photonic crystal structures [42,62].

It should be stressed that the structure considered here allows for considerable size of the topological gap, the property that is important for applications [63]. According to the spectrum presented in Fig. 2, one finds that for $r = 0.4a$ the bandgap opening in the spectrum constitutes up to $42.8\%$ according to $\text{ratio} = (\varepsilon_{\text{top}} - \varepsilon_{\text{bottom}})/\varepsilon_{\text{middle}}$, where $\varepsilon_{\text{top}}$, $\varepsilon_{\text{middle}}$ and $\varepsilon_{\text{bottom}}$ correspond to the top edge, middle and bottom edge of the bandgap. Considering that there are both corner and edge states in the bandgap, the width of the gap (where corner state is located) between the edge states and top of the bulk band is about $26.51\%$ of the whole band width. We believe that the ratio can be improved if the value of $r$ is decreased below $0.4a$. However, there are some limitations, because for too small values of $r$ the neighboring sites in the array may start fusing. Large size of the topological gap is beneficial for topological protection of the lasing regime.

The important property of topological lasers is that lasing in this structure remains stable as long as variations of eigenvalues of the modes that it supports due to inevitable disorder upon fabrication of the structure does not lead to closure of the gap, so that corner mode persists and experiences largest gain, when pump is provided in the corner waveguide. In **Supplemental Materials** we introduce various types of perturbations (most of them break spatial symmetry of the system [64]) into Kagome array, including random variations of waveguide depths/ positions in the bulk, at the edge, or only in the corner sites. For all these types of perturbations robustness of the corner state was confirmed. The conclusion is that when disorder is provided in depths of the bulk and edge waveguides of the array, but not in the corner ones, the gap and corner mode residing in the gap persist even for $10\%$ perturbations. When disorder is present also in the corner waveguide, its impact on corner mode is stronger and may lead to shift of the eigenvalue of the corner mode, but for moderate disorder amplitudes, corner states also persist in the gap and do not couple to bulk modes. It should be stressed that for latter type of disorder the energies of the corner states start overlapping with continuous band only when the actual strength of disorder $\delta \cdot p_{\text{re}}$ [here we introduce disorder by changing the depth of the waveguide with indices $n, m$ to $p_{\text{re}}(1 + \delta_{n,m})$, where $\delta_{n,m}$ is a random number uniformly distributed within the interval $[-\delta, +\delta]$], is comparable with the width of the topological gap, i.e. when disorder is not small. Shifting the top-right corner site of the structure along the direction parallel/orthogonal to the line connecting the top-right and bottom-left corners of rhombus also does not lead to destruction of the corner state and demonstrates its robustness. A table that illustrates all above mentioned perturbations, examples of corresponding robust corner states, and spectra of all modes of the system (illustrating that no new modes appeared in the gap) can be found in the Supplemental Materials. Even in the presence of disorder, when localized gain is provided in the corner channel, corner modes do lase above lowest threshold, so in this sense the robustness of lasing operation with respect to disorder is obvious. It should be mentioned that typical uncertainties in the fabrication process of the structure strongly depend on the particular technology, for example for direct laser writing of waveguides, they usually do not exceed 2% [65].

To support the claim of more robust behavior of the topological system, we also compare the behavior of the same platform in topological and nontopological regimes. Thus, in Supplementary Materials we show spectrum of trivial array (i.e. the array with $r = a/2$) and examples of corresponding modes that are all delocalized. We found that when gain is provided in the corner of trivial Kagome array with $r = a/2$, the threshold for lasing is substantially higher than that in the topological phase, at $r = 0.4a$. The family of nonlinear lasing modes now bifurcates from one of delocalized linear modes, since there is no gap in the spectrum. Due to nonlinear competition of several bulk modes, they may mix in lasing state. This is manifested in change of slope of the dependence of the maximal wave amplitude and nonlinear energy shift on gain parameter.

There is ongoing debate on topological characterization of modes in Kagome arrays [66,67] with many evidences of topological behavior of this system reported experimentally. The emergence of corner modes in the spectrum of our continuous system (that automatically takes into account long-range coupling) was always correlated with the appearance of nontrivial bulk polarization calculated in the frames of simplified discrete model. While this may not be the case for multilayer topological systems, here we would like to emphasize that Kagome array under the action of perturbations behaves very similarly to the well-established topological Su-Schrieffer-Heeger (SSH) model, and at the same time it remains truly two-dimensional system. The comparison of the robustness of corner modes of Kagome array and topological modes of the SSH model under the action of different perturbations can be found in Supplemental Materials. We find that in both models, energies of the topological corner states may shift into band, but only when perturbations are added into corner (or edge in the case of SSH chain) site and when their amplitude is comparable with the width of the whole topological gap. As confirmed by calculation of bulk polarization, this supports the conclusion about topological nature of corner states in Kagome array.

## 3 Conclusions

In summary, we have proposed corner state laser based on Kagome waveguide array with rhombic configuration. Stable corner lasing states with low threshold bifurcate from the topological corner modes, under the balance between diffraction, focusing nonlinearity, uniform loss, two-photon absorption, and gain provided in selected corner of the structure, supporting topological states. Applying gain in other corners of the array allows achieving lasing in edge states or bulk modes, but above much higher lasing thresholds. Excitation of topological and nontopological nonlinear modes from noisy inputs is illustrated too, and it is found that formation of nontopological modes requires much longer propagation distances and they appear to be weakly localized, in contrast to strongly localized nonlinear corner lasing modes. This work paves the way to realization of 0D topological corner lasers based on HOTIs and may inspire research of topological transitions in similar dissipative systems.

## Supplemental Materials

See supplemental materials for the details about the bulk polarizations, robustness analysis on the corner state, and other issues mentioned in the main text.

## Acknowledgments

This work was supported by the Guangdong Basic and Applied Basic Research Foundation (2018A0303130057), National Natural Science Foundation of China (12074308, U1537210), RFBR and DFG according to the research Project No. 18-502-12080, and the Fundamental Research Funds for the Central Universities (xzy012019038, xzy022019076). H.Z. and Y.Q.Z. acknowledge the computational resources provided by the HPC platform of Xi'an Jiaotong University.

## Data Availability

The data that support the findings of this study are available from the corresponding author upon reasonable request.

# Supplemental Materials

## 1. Bulk polarization

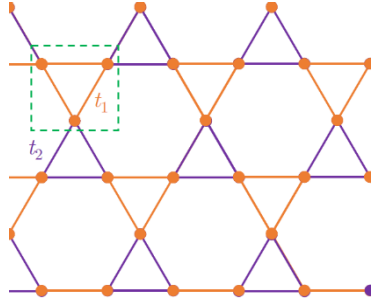

Fig. S1 (a) Schematic diagram of kagome array. The orange and purple lines indicate nearest-neighbor coupling strengths $t_1$ (downward triangles) and $t_2$ (upward triangles). The unit cell is marked by the green squares.

We consider the Hamiltonian on a kagome array with parallelogram structure as:

$$H = \begin{pmatrix} 0 & t_1 + t_2 e^{i\mathbf{k}\cdot\mathbf{e}_1} & t_1 + t_2 e^{-i\mathbf{k}\cdot\mathbf{e}_2} \\ t_1 + t_2 e^{-i\mathbf{k}\cdot\mathbf{e}_1} & 0 & t_1 + t_2 e^{i\mathbf{k}\cdot\mathbf{e}_3} \\ t_1 + t_2 e^{i\mathbf{k}\cdot\mathbf{e}_2} & t_1 + t_2 e^{-i\mathbf{k}\cdot\mathbf{e}_3} & 0 \end{pmatrix} \quad (S1)$$

where $\mathbf{e}_1 = (1/2, -\sqrt{3}/2)a$, $\mathbf{e}_2 = (-1,0)a$, $\mathbf{e}_3 = (1/2, \sqrt{3}/2)a$, $\mathbf{k} = (k_x, k_y)$, $t_1$ and $t_2$ are intra- and inter- nearest-neighbor coupling strengths, respectively. The unit cell is defined by the triangle marked by the dashed green square in Fig. S1. The coupling strengths can be manipulated by the distance between two neighboring waveguides (please see Fig. 2 in the main text). The relation is:

$$\begin{cases} t_1 = t_2, & \text{if } r = 0.5a \\ t_1 > t_2, & \text{if } r > 0.5a \\ t_1 < t_2, & \text{if } r < 0.5a \end{cases} \quad (S2)$$

For the case with $t_1 = t_2$, one could see there are two dispersive bands with a flat band, and there is no bandgap between two dispersive bands. While if $t_1 \neq t_2$, there exists a bandgap between the first two bands. In addition, the next nearest intra- and inter- unit cell distances can control the topological property of the band. For $t_1 > t_2$, the bandgap is topologically trivial, whereas for $t_1 < t_2$, the bandgap is topologically nontrivial. In other words, the system is topologically nontrivial if $r < 0.5a$. The topological properties are characterized by the bulk polarization, that is defined as:

$$p_j = \frac{i}{S} \int_S d^2\mathbf{k} \langle u_m(\mathbf{k}) | \partial_{k_j} u_n(\mathbf{k}) \rangle, \quad (S3)$$

in which, $S$ is the size of first Brillouin zone, $u_m(\mathbf{k})$ is the periodical Bloch function for the $m$th band that can be obtained by solving the Hamiltonian in Eq. (S1). Since the first Brillouin zone of the kagome array is not square, it is better to make a coordinate transform, that leads to its transformation into square zone, as shown Fig. S2. So, the bulk polarizations will change from $(p_1, p_2)$ to $(p'_x, p'_y)$ in the new coordinate system.

One obtains in this system, that for $t_1 > t_2$ ($r > 0.5a$), the topological indices $(p'_x, p'_y) = (0,0)$, that means that the system is topologically trivial. While for $t_1 < t_2$ ($r < 0.5a$), $(p'_x, p'_y) = (1/3, 1/3)$, that corresponds to the topologically nontrivial phase.

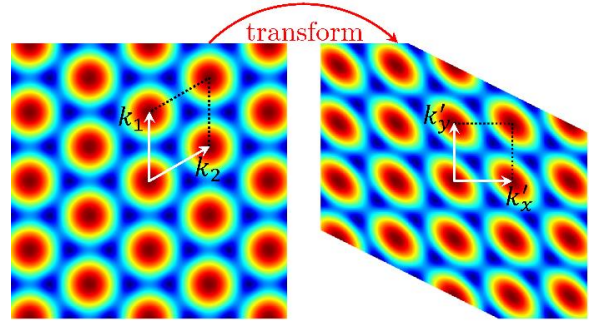

Fig. S2. Transformation of the band structure. The first Brillouin zone with two lattice vectors $k_1$ and $k_2$ (left) is transformed into a square first Brillouin zone with lattice vectors $k'_x$ and $k'_y$.

## 2. Lasing in the presence of disorder
### 2.1 Robustness of the corner states

To prove robustness of the corner states in kagome arrays and to illustrate how disorder affects their energies we added disorder (breaking all geometrical symmetries) to the bulk sites (except for the edge and corner sites), to the edge sites (except for the corner ones), and only to the corner sites of the structure, respectively. Thus, the depth of the site with index $(n, m)$ was taken as $p_{\text{re}}(1 + \delta_{n,m})$, where $\delta_{n,m}$ is the random number uniformly distributed within the interval $[-\delta, +\delta]$, while $p_{\text{re}}$ is the refractive index in the unperturbed structure. The transformation of the spectra of the modes with increase of the strength of disorder is displayed in Fig. S3. The red dots represent the corner state. One observes that the corner state persists in all cases and that is it quite robust. Only for the disorder added into corner sites the energies start to shift, but they start overlapping with the continuous band only when the strength of disorder (that can be estimated as $\delta \cdot p_{\text{re}}$) is comparable with the width of the topological gap, i.e. when disorder is not small.

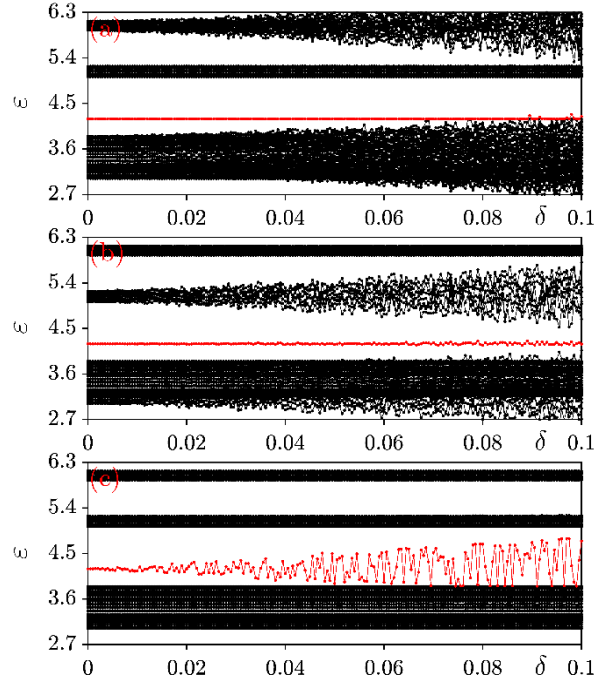

Fig. S3. (a) Variation in the spectrum of eigenmodes in rhombic kagome configuration with increase of disorder strength, when disorder is introduced into the bulk waveguides, but not into corner or edges ones. (b) Variation in the spectrum of eigenmodes in rhombic configuration with increase of disorder strength, when disorder is introduced only into edge waveguides. (c) Variation in the spectrum of eigenmodes in rhombic configuration with increase of disorder strength, when disorder is introduced only into corner waveguide. Red dots in all plots correspond to the corner states. Here $r = 0.4a$ and $\delta$ represents the strength of the disorder.

We also shifted the top-right corner site of the structure along the direction parallel/orthogonal to the line connecting the top-right and bottom-left corners of rhombus to check the robustness of the corner state. We show corresponding variation in the spectrum of the array in Fig. S4. In Fig. S4(a), the shift is parallel to the above mentioned line, while in Fig. S4(b) it is orthogonal to that line. Red dots indicate the corner state. Clearly, the corner state is also robust to such a perturbation.

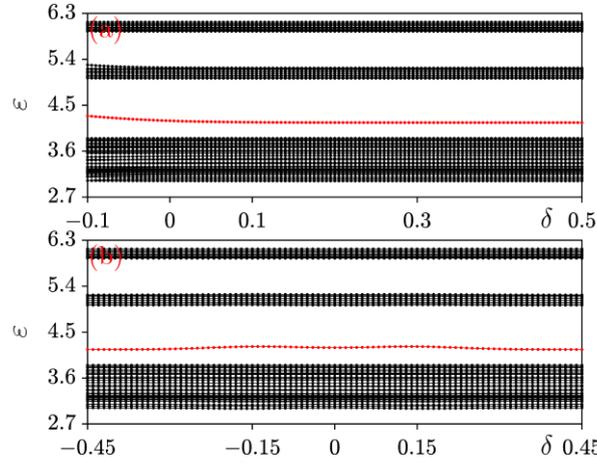

Fig. S4. (a) Variation in the spectrum of eigenmodes in rhombic configuration upon motion of the top-right corner site along (a) and perpendicularly (b) to the direction parallel to the line connecting the top-right and bottom-left corners. $\delta$ represents the displacement.

In the table on the next page, we accumulate all information about the types of disorder considered for kagome array, present examples of the profiles of the corner states, and show spectra of eigenmodes to stress that it is not destroyed by added perturbations.

| *Perturbation type* | *Corner state profile* | *Spectrum of eigenmodes* |
|---|---|---|
| 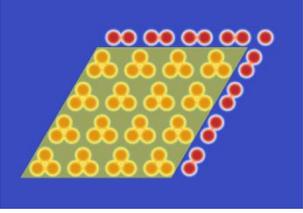 Disorder is added to the shaded sites (bulk sites) with amplitude up to $\delta = 0.04$ | 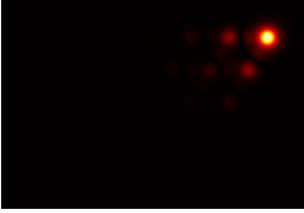 | 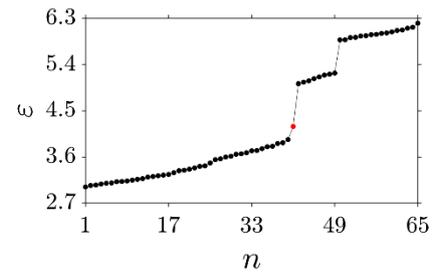 |
| 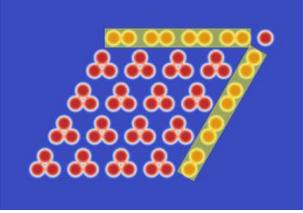 Disorder is added to the shaded sites (edge sites) with amplitude up to $\delta = 0.04$ | 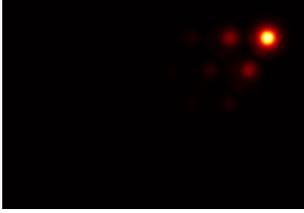 | 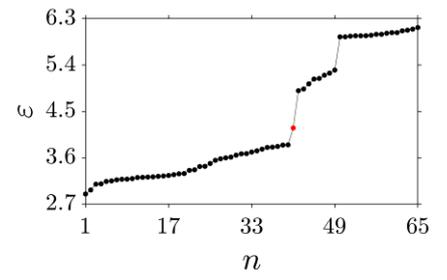 |
| 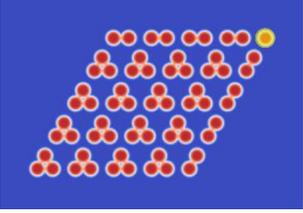 Disorder is added to the corner site with the amplitude up to $\delta = 0.04$ | 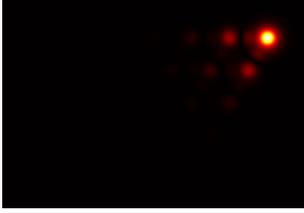 | 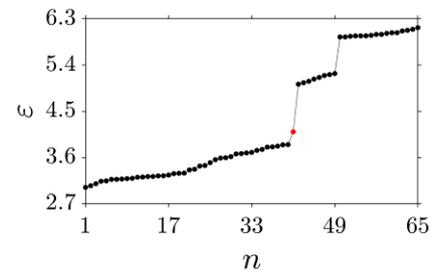 |
| 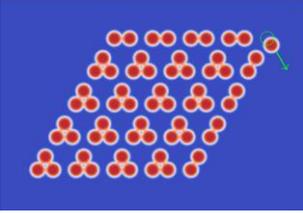 Corner site is shifted along the direction indicated the by the arrow by $\delta = 0.1$. Green circle represents the original corner site. | 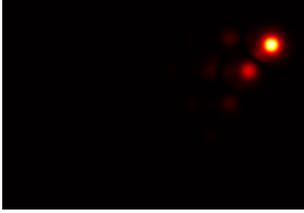 | 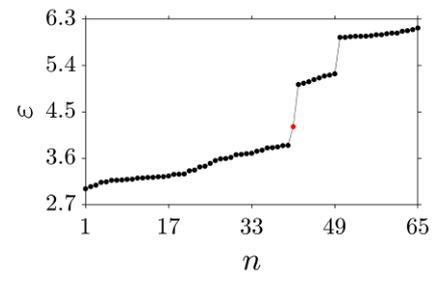 |
| 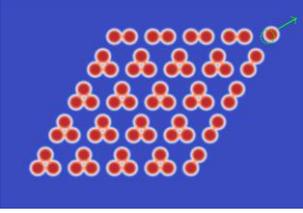 Corner site is shifted along the direction indicated the by the arrow by $\delta = 0.1$. Green circle represents the original corner site. | 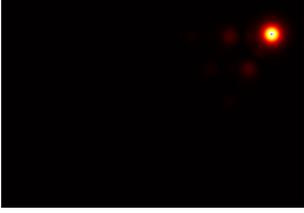 | 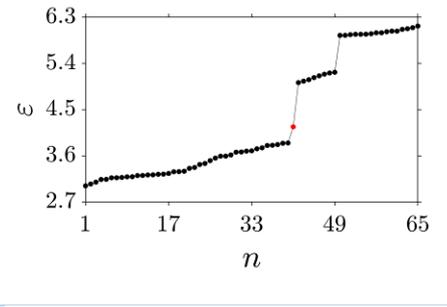 |
| 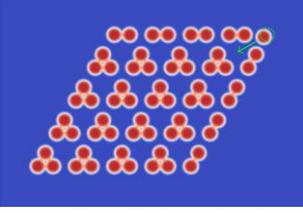 Corner site is shifted along the direction indicated the by the arrow by $\delta = 0.1$. Green circle represents the original corner site. | 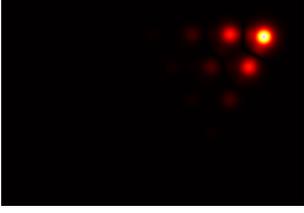 | 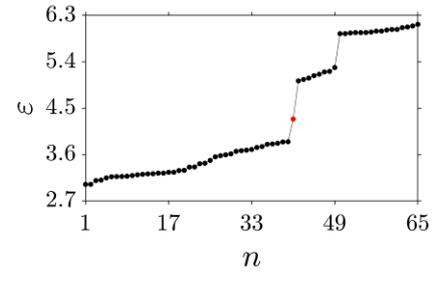 |

Notice that in all cases considered, corner states in the gap of kagome array persist and, importantly, no new corner modes are generated by the disorder or waveguide shifts (i.e. all states except for corner one, marked by the red dot, remain extended, edge states remain extended along the edge). The behavior is qualitatively similar to that of the well-established topological SSH model as illustrated below.

## 2.2 Dynamics of the lasing state with disorder in the corner site

One of the main advantages of topological lasers is the resistance to disorder in the underlying laser structure. To investigate, how perturbations affect lasing properties of our structure in nonlinear regime, we first increased and decreased refractive index of the corner waveguide, as shown in Figs. S5(a-d). Clearly, the stable lasing regime was not affected and amplitude of the lasing mode experiences only small variations. In Fig. S5(a) we show the amplitude of the lasing mode versus propagation distance, where the yellow line corresponds to the case, when refractive index was increased by 10%, the red line corresponds to evolution of slightly perturbed lasing mode when all waveguides are identical, and the blue line corresponds to the case, when refractive index of the corner waveguide was decreased by 10%. The profiles of lasing modes for the three cases at $z = 1000$ are shown in Figs. S5(b-d). Second, we added disorder into all waveguides except for the corner one by assuming that the refractive index in the bulk and edge waveguides is given by $p_{re}(1 + \delta_{n,m})$, where $\delta_{n,m}$ is the random number. The results are displayed in Figs. S5(e-g). In Fig. S5(e), the red line corresponds to the case with $\delta_{n,m}$ uniformly distributed within the interval $[0, 0.1]$, and the blue line corresponds to the case with $\delta_{n,m}$ uniformly distributed within the interval $[-0.1, 0]$. Again, stable lasing regime was not affected and amplitude of the lasing mode experiences only small oscillations, quickly adjusting to new asymptotic value. The lasing modes corresponding to the two cases at $z = 1000$ are exhibited in Figs. S5(f) and S5(g). We thus conclude that the presence of disorder does not destroy lasing regime, leading to only small variations of the amplitude of corner lasing states.

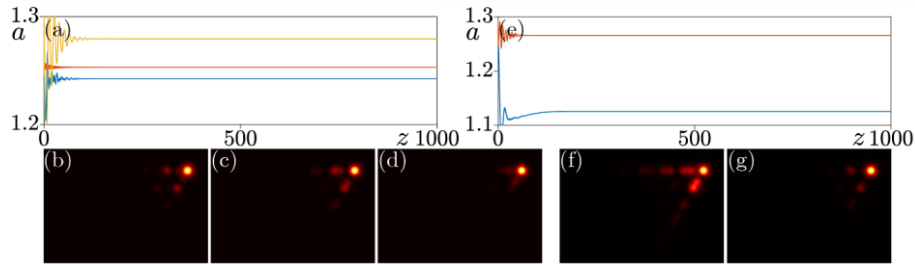

Fig. S5. (a) Amplitude of the lasing mode $a$ versus propagation distance $z$. Yellow, red, and blue lines correspond to modification of the refractive index of the corner waveguide by $+10\%$, $0\%$, and $-10\%$, respectively. (b)-(d) Profiles of the lasing modes at $z = 1000$, corresponding to the blue, red and yellow cases, respectively. (e) Amplitude of the lasing mode $a$ versus propagation distance $z$. Red and blue lines correspond to the amplitude of the random noise superposed to the sites except the corner waveguide to $+10\%$ and $-10\%$, respectively. (f,g) Profiles of the lasing modes at $z = 1000$, corresponding to the red and blue cases, respectively.

## 3. Comparison between the SSH model and the kagome array

In this section we illustrate that topological corner states in two-dimensional kagome array show some similarities with behavior of the edge states in well-established topological Su-Schrieffer-Heeger (SSH) model [the array of two-dimensional waveguides with exactly the same parameters as for kagome array, but arranged into line SSH configuration, Fig. S6(a)]. In particular, our simulations confirm very similar behavior of modes of these two topological systems under the action of perturbations/disorder.

We show the spectrum of the SSH model and discuss the robustness of its edge states in Fig. S6. As shown in Fig. S6(b), the separation $r$ indeed qualitatively affects the spectrum. When $r \neq 0.5a$, there is a band gap, and the edge states (indicated by the red curve in the spectrum) only exists in the region with $r < 0.5a$. The impact of disorder (that breaks the symmetry of the array) added into bulk waveguides is illustrated in Fig. S6(c). This type of disorder practically does not affect the energies of the edge states. When disorder is added into boundary waveguides (it also breaks the symmetry of the array, because it typically has different amplitudes in two boundary waveguides) the energies of the edge state do experience variation [Fig. S6(d)], but they start overlapping with bulk bands only when disorder parameter $\delta \sim 0.08$. This disorder is not small – for $p_{\text{re}} = 10$ used here one estimates fluctuations in the depth of the waveguides as $\delta \cdot p_{\text{re}}$ and for $\delta \sim 0.08$ this number is comparable with the width of the entire topological gap. Therefore, the energies of the edge states start overlapping with bulk bands only when disorder strength becomes comparable with gap width.

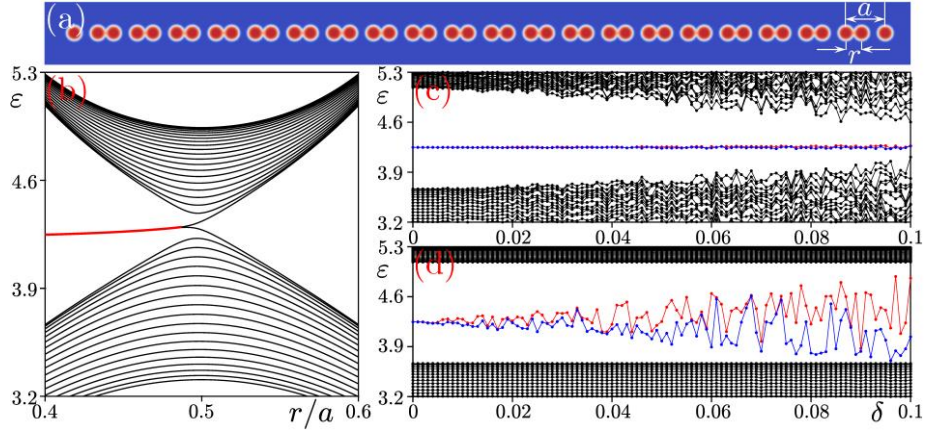

Fig. S6. (a) A SSH chain with $r = 0.4a$. Other parameters are same as those adopted in the manuscript. (b) Spectrum $\varepsilon$ of the SSH chain versus $r$, in which the red curve indicating the boundary state. (c) Spectrum $\varepsilon$ of the SSH chain with $r = 0.4a$ and random noise added to the sites except the two boundary sites through $p_{\text{re}}(1 + \delta_{\text{m,n}})$ where $\delta_{\text{n,m}}$ is a random number. (d) Setup is as (c), but with random noise added to the two boundary sites only.

For comparison, in Fig. S7 we display the spectrum of the triangular breathing kagome array versus $r$. Clearly, one finds that the spectra in Fig. S6(b) and S7(b) are quite similar, even though effective dimensionalities of two systems are different – variation in separation $r$ results in the appearance of the topological states. Notice that in both cases they appear after closing and reopening of the gap, which is a typical behavior for topological systems.

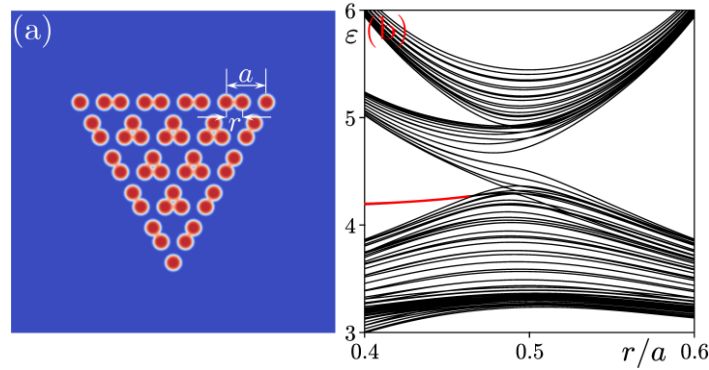

Fig. S7. (a) A triangular kagome array with $r = 0.4a$. Other parameters are same as those adopted in the manuscript. (b) Spectrum $\varepsilon$ of the triangular kagome array versus $r$, in which red curve indicates corner states.

The results in Sections 2 and 3 confirm topological nature of the model and the fact that lasing in this system is topologically protected.

## 4. Lasing threshold and nonlinear lasing mode family in triangular configuration

Following the same procedure adopted in the main text, we introduce gain into the triangular configuration (bottom corner site) to check the lasing threshold. We obtain the spectrum of the triangular configuration with gain, as shown in Fig. S8(a). Similarly to rhombic configuration, the lasing threshold in the triangular configuration is $p_{\text{im}}^{\text{th}} = 0.078$.

We also calculated the family of nonlinear lasing modes corresponding to the case with $\alpha = 0.4$, and the result is displayed in Fig. S8(b). Again the result is the same as that for the rhombic configuration [please see the right curve in Fig. 5(a) in the main text].

According to the results in this section and section 3.1 in main text, one finds that the results remain identical for both configurations. However, rhombic configuration is advantageous because there is no degeneracy of corner states in that structure.

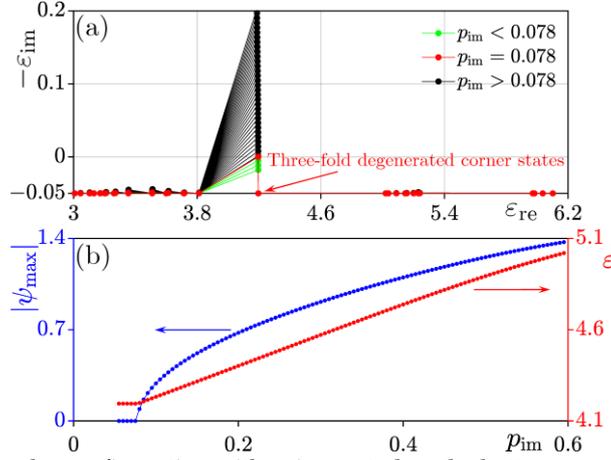

Fig. S8. (a) Spectrum of the triangular configuration with gain exerted on the bottom corner site. (b) Nonlinear lasing mode family with $\alpha = 0.4$.

## 5. Dependence of lasing threshold on the coefficient $r$

As demonstrated in section 1, corner states emerge if $r < 0.5a$. In our system the value of $r$ cannot be too small, because neighboring sites in the array should not overlap. Therefore, we change the value of $r$ in the range $0.39 \leq r < 0.5$ to check how this parameter affects the laser characteristics.

Here, we adopt the triangular configuration, but the conclusion is also feasible to the rhombic configuration. We first check the lasing threshold by changing $r$, and the result is shown by the black curve in Fig. S9. One finds that with increase of $r$, the lasing threshold $p_{\text{im}}^{\text{th}}$ gradually increases and that it approaches some constant at small values of $r$.

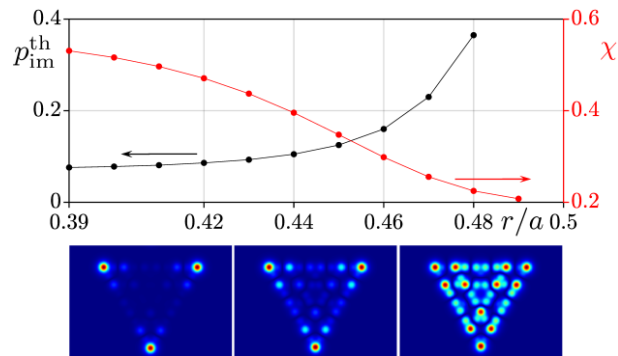

Fig. S9. Lasing threshold (black curve) and form-factor (inversely proportional to width) of the linear corner state (red curve) versus $r$. The bottom panels show the amplitudes of the linear corner modes (from left to right) with $r/a = 0.39$, $0.44$ and $0.49$, respectively.

In addition to the lasing threshold, the value of $r$ also substantially affects the localization of the corner state and thus the profile of the lasing mode. In Fig. S9, we also display the form factor (inverse width) of the linear corner state calculated in accordance with Eq. (S4), see the red curve. Clearly, the smaller the value of $r$, the better the localization. In the bottom panels, we exhibit the amplitude profiles of the linear corner states corresponding to $r/a = 0.39,\ 0.44$ and $0.49$, respectively.

$$\chi = \left(\iint |\psi|^4 dx dy\right)^{1/2} \left(\iint |\psi|^2 dx dy\right)^{-1}. \tag{S4}$$

### 6. Lasing in topologically trivial rhombic configuration

The spectrum of eigenmodes of the kagome array with rhombic configuration in trivial phase at $r = 0.5a$ is displayed in Fig. S10. In contrast to the spectrum of topologically nontrivial structure with $r = 0.4a$ in Fig. 2 in the main text, there is no gap and no corner states in the spectrum. The examples of states corresponding to the red dots from different parts of the spectrum are provided in the bottom row of Fig. S10. As shown below, these states will experience strongest amplification above the lasing threshold, but they are not corner states and they strongly expand into the bulk of the array.

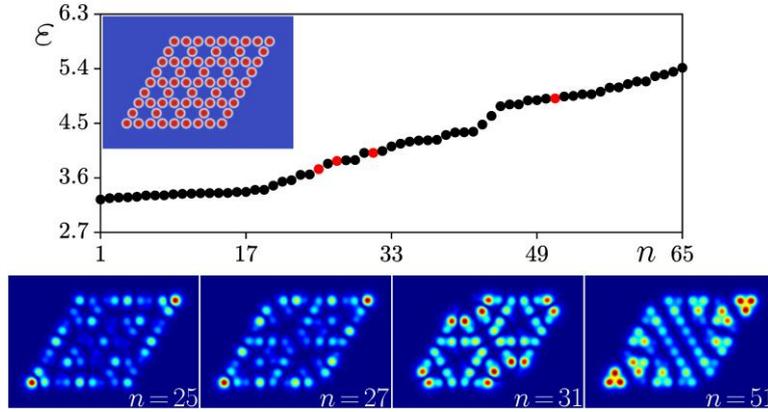

Fig. S10. Spectrum of the rhombic configuration in nontopological regime with $r = 0.5a$. The configuration is displayed in the inset. Corresponding to the red dots in the spectrum, the states are displayed in the bottom panels with the number shown in the bottom-right corner of each panel.

We next introduce gain into the top-right corner site to check lasing thresholds for different modes. The imaginary parts of the eigenvalues for different modes are plotted in Fig. S11(a) for different gain amplitudes. We found that lasing threshold in topologically trivial structure is $p_{\text{im}}^{\text{th}} = 0.42$ and it substantially exceeds lasing threshold $p_{\text{im}}^{\text{th}} = 0.078$ for topological structure. Since only bulk states experience now preferential amplification (and there are several of such states), lasing occurs in bulk modes and it is characterized by substantially longer transition to steady-state regime.

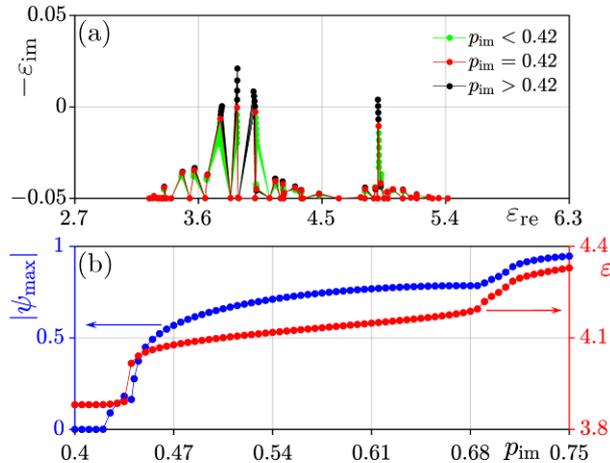

Fig. S11. (a) Spectrum of the modes in trivial kagome configuration (the configuration in Fig. S10) with gain exerted on the top-right corner site. There are four states that start lasing first and they are depicted in the bottom row of Fig. S10. Among these four states, the one with $n = 27$ lases first. (b) Nonlinear lasing mode family of the $n = 27$ mode with $\alpha = 0.4$. At $p_{\text{im}} = 0.42$, the nonlinear lasing mode bifurcates from linear one.

We also calculated the family of nonlinear lasing modes bifurcating from the $n = 27$ mode at $\alpha = 0.4$, and the result is displayed in Fig. S11(b). Since there is no gap in the spectrum in Fig. S10, the bulk modes mix up with each other. This is manifested in change of slope of the dependence of the maximal amplitude and nonlinear energy shift on gain parameter visible in Fig. S11(b) [these dependencies also differ substantially from those for topological modes, see Fig. S8(b) and the right curve in Fig. 5(a) in the main text].

## 7. Comparison of lasing in a single waveguide and topological Kagome array

As shown in Fig. S12 with dependencies of peak amplitude $a$, power $P$, and mode energy $\varepsilon$ on gain amplitude for Kagome array and isolated waveguide, there is no fundamental difference in lasing efficiency for them. In the case of Kagome array the structure of the lasing mode strongly depends on the dimerization parameter $r$. The field of this mode is localized mostly on one of the sublattices forming Kagome array and it changes its sign between unit cells, that results in staggered tails, as it is obvious from profile of the corner mode ($n = 41$) from Fig. 2 in the main text. Thus, the presence of the array does allow to control the spatial extent of the lasing pattern and is therefore very important (thus, at $r$ values close to $a/2$ lasing will occur in the mode strongly penetrating into the depth of array, even though gain is provided only in one waveguide). Notice that in the case of the array coupling to bulk modes appear at sufficiently high gain amplitudes, the feature that is absent for single waveguide. It should be also mentioned that in the limit of strong dimerization (corresponding to small shifts $r$), when corner waveguide detaches more and more from the array, the field strongly localizes on corner waveguide and corresponding lasing characteristics naturally approach closely those for the isolated waveguide. As concerns resistance to disorder, it is also present in HOTIs, the energy of the corner mode remains in the topological gap and no new localized modes appear in this gap as long as disorder does not destroy it.

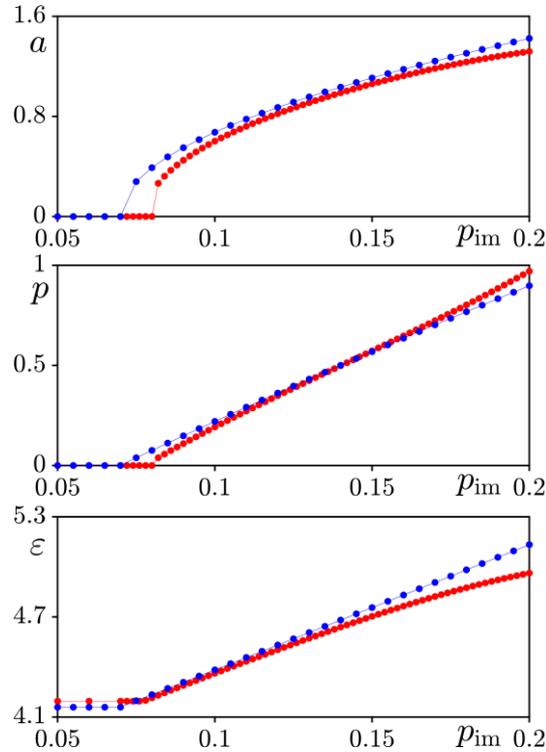

Fig. S12. Peak amplitude $a$, power $P$ and energy $\varepsilon$ of the lasing mode for topological corner waveguide (red curve) and single waveguide (blue curve). The parameters are same as those used for Fig. 5(a) in the main text. Here, we choose $\alpha = 0.1$.